\documentclass[9pt,twocolumn,twoside]{optica}
\setboolean{shortarticle}{false}
\setboolean{minireview}{false}

\DeclareMathOperator*{\argmin}{arg\,min}
\usepackage{float}
\usepackage{xcolor}
\usepackage{siunitx}

\newcommand\mytilde{\raise.17ex\hbox{$\scriptstyle\sim$}}
\newcommand\sensorpixelx{448}
\newcommand\sensorpixely{320}

\title{Spectral DiffuserCam: lensless snapshot hyperspectral imaging with a spectral filter array}
\author[1,$\dagger$,*]{Kristina Monakhova}
\author[2,$\dagger$,*]{Kyrollos Yanny}
\author[1]{Neerja Aggarwal}
\author[1,2]{Laura Waller}

\affil[1]{Department of Electrical Engineering \& Computer Sciences, University of California, Berkeley, CA, 94720, USA}
\affil[2]{UCB/UCSF Joint Graduate Program in Bioengineering, University of California, Berkeley, CA,
94720, USA}

\affil[*]{Corresponding authors: monakhova@berkeley.edu, kyrollosyanny@gmail.com}

\affil[$\dagger$]{These authors contributed equally to this work}
\begin{abstract}
Hyperspectral imaging is useful for applications ranging from medical diagnostics to agricultural crop monitoring; however, traditional scanning hyperspectral imagers are prohibitively slow and expensive for widespread adoption. Snapshot techniques exist but are often confined to bulky benchtop setups or have low spatio-spectral resolution. In this paper, we propose a novel, compact, and inexpensive computational camera for snapshot hyperspectral imaging. Our system consists of a tiled spectral filter array placed directly on the image sensor and a diffuser placed close to the sensor. Each point in the world maps to a unique pseudorandom pattern on the spectral filter array, which encodes multiplexed spatio-spectral information. By solving a sparsity-constrained inverse problem, we recover the hyperspectral volume with sub-superpixel resolution. Our hyperspectral imaging framework is flexible and can be designed with contiguous or non-contiguous spectral filters that can be chosen for a given application. We provide theory for system design, demonstrate a prototype device, and present experimental results with high spatio-spectral resolution.
\end{abstract}

\setboolean{displaycopyright}{true}

\begin{document}

\maketitle

\section{Introduction}
Hyperspectral imaging systems aim to capture a 3D spatio-spectral cube containing spectral information for each spatial location. This enables the detection and classification of different material properties through spectral fingerprints, which cannot be seen with an RGB camera alone. Hyperspectral imaging has been shown to be useful for a variety of applications, from agricultural crop monitoring to medical diagnostics, microscopy, and food quality analysis~\cite{delalieux2009hyperspectral,kester2011real,lu2014medical, sun2010hyperspectral, gowen2007hyperspectral, akbari2012hyperspectral, lu2014spectral, orth2015gigapixel, huang2015development, bacon2004miniature}. Despite the potential utility, commercial hyperspectral cameras range from \$25,000 - \$100,000 (at the time of publication of this paper). This high price point and the large size have limited the widespread use of hyperspectral imagers.

Traditional hyperspectral imagers rely on scanning either the spectral or spatial dimension of the hyperspectral cube with spectral filters or line-scanning~\cite{green1998imaging, gat2000imaging, zhang2016novel}. These methods can be slow and generally require precise moving parts, increasing the camera complexity. More recently, snapshot techniques have emerged, enabling capture of the full hyperspectral data cube in a single shot. Some snapshot methods trade-off spatial resolution for spectral resolution by using a color filter array or splitting up the camera's field-of-view (FOV). Computational imaging approaches can circumvent this trade-off by spatio-spectrally encoding the incoming light, then solving a compressive sensing inverse problem to recover the spectral cube~\cite{wagadarikar2008single}, assuming some structure in the scene. These systems are typically table-top instruments with bulky relay lenses, prisms, or diffractive elements, suitable for laboratory experiments, but not the real world.  Recently, several compact snapshot hyperspectral imagers have been demonstrated that encode spatio-spectral information with a single optic, enabling a practical form factor~\cite{sahoo2017single, french2017speckle,jeon2019compact}. Using a single optic to control both the spectral and spatial resolution, they are generally constrained to measuring contiguous spectral bins within a given spectral band. 

\begin{figure*}[thb]
	\centering\includegraphics[width= \linewidth]{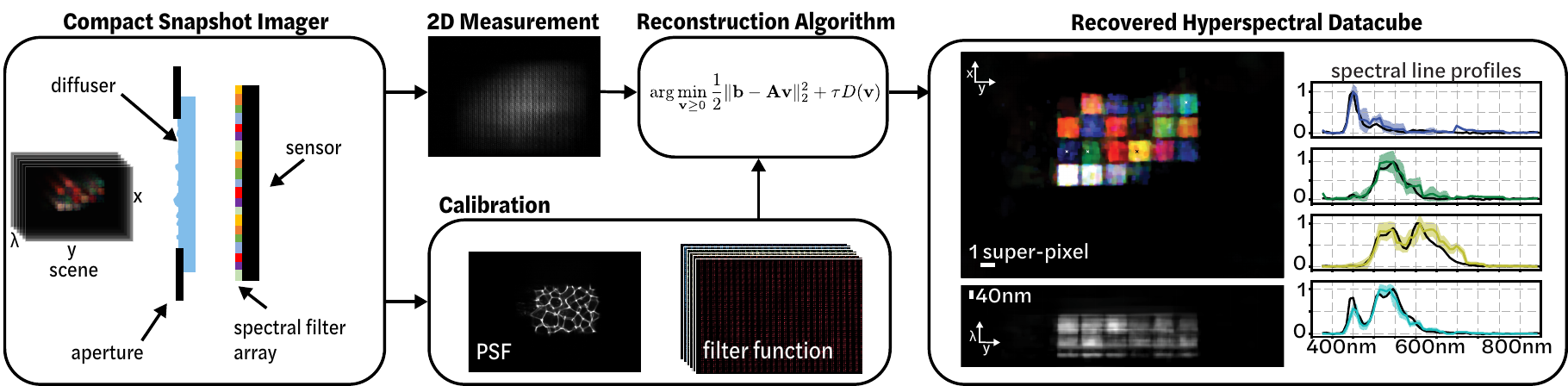}
 \caption{Overview of the Spectral DiffuserCam imaging pipeline, which reconstructs a hyperspectral datacube from a single-shot 2D measurement. The system consists of a diffuser and spectral filter array bonded to an image sensor. A one-time calibration procedure measures the point spread function (PSF) and filter function. Images are reconstructed using a non-linear inverse problem solver with a sparsity prior. The result is a 3D hyperspectral cube with 64 channels of spectral information for each of \sensorpixelx$\times$\sensorpixely{} spatial points, generated from a 2D sensor measurement that is \sensorpixelx$\times$\sensorpixely{} pixels.}
 \label{fig:overview}
\end{figure*}

Here, we propose a new encoding scheme that takes advantage of recent advances in patterned thin film spectral filters~\cite{saxe2018advances}, and lensless imaging, to achieve high-resolution snapshot hyperspectral imaging in a small form factor. Our system consists of a tiled spectral filter array placed directly onto the sensor and a randomizing phase mask (i.e. diffuser) placed a small distance away from the sensor, as in the DiffuserCam architecture~\cite{antipa2018diffusercam}. The diffuser spatially multiplexes the incoming light, such that each spatial point in the world maps to many pixels on the camera. The spectral filter array then spectrally encodes the incoming light via a structured erasure function. The multiplexing effect of the diffuser allows recovery of scene information from a subset of sensor pixels, so we are able to recover the full spatio-spectral cube without the loss in resolution that would result from using a non-multiplexing optic, such as a lens. 
%the spectral filter array would result in holes in the measurement or the need for lower spatial resolution, the pseudo-random encoding caused by the diffuser allows us to solve a compressive sensing inverse problem to recover higher spatial and spectral resolution than would be possible with a lens-based system. 
%In addition, the diffuser enables us to use a regularly spaced spectral filter array.

Our encoding scheme enables hyperspectral recovery in a compact and inexpensive form factor. The spectral filter array can be manufactured directly on the sensor, costing under \$5 for both the diffuser and the filter array at scale. A key advantage of our system over previous compact snapshot hyperspectral imagers is that it decouples the spectral and spatial responses, enabling a flexible design in which either contiguous or non-contiguous spectral filters with user-selected bandwidths can be chosen. Given some conditions on scene sparsity and the diffuser randomness, the spectral sampling is determined by the spectral filters and the spatial resolution is determined by the autocorrelation of the diffuser response. This should find use in task-specific/classification applications~\cite{saragadam2020programmable, chao2007hyperspectral, levenson2008multiplexing, hennessy2020hyperspectral}, where one may wish to tailor the spectral sampling to the application by measuring multiple non-contiguous spectral bands, or have higher-resolution spectral sampling for certain bands. 

We present theory for our system, simulations to motivate the need for a diffuser, and experimental results from a prototype system. The main contributions of our paper are:
\begin{enumerate}[noitemsep,nolistsep]
 \item A novel framework for snapshot hyperspectral imaging that combines compressive sensing with spectral filter arrays, enabling compact and inexpensive hyperspectral imaging.
 \item Theory and simulations analyzing the system's spatio-spectral resolution for objects with varying complexity. 
 \item A prototype device demonstrating snapshot hyperspectral recovery on real data from natural scenes.
\end{enumerate}

\section{Related Work}
\subsection{Snapshot Hyperspectral Imaging}
There have been a variety of snapshot hyperspectral imaging techniques proposed and evaluated over the past decades. Most approaches can be categorized into the following groups: spectral filter array methods, coded aperture methods, speckle-based methods, and dispersion-based methods. 

\textbf{Spectral filter array methods} use tiled spectral filter arrays on the sensor to recover the spectral channels of interest~\cite{lapray2014multispectral}. These methods can be viewed as an extension of Bayer filters for RGB imaging, since each `super-pixel' in the tiled array has a grid of spectral filters. As the number of filters increases, the spectral resolution increases and the spatial resolution decreases. For instance, with an 8$\times$8 filter array (64 spectral channels), the spatial resolution is 8$\times$ worse in each direction than that of the camera sensor. Demosaicing methods have been proposed to improve upon this in post-processing; however, they rely on intelligently guessing information that is not recorded by the sensor~\cite{mihoubi2017multispectral}. Recently, photonic crystal slabs have been demonstrated for compact spectroscopy based on random spectral responses (as opposed to traditional passband responses) and extended to hyperspectral imaging through the tiling of the photonic crystal slab pixels~\cite{wang2014spectral, wang2019single}. While these methods have high spectral accuracy, they have only been demonstrated in a 10$\times$10 spatial pixel configuration. Our system uses a spectral filter array, but combines it with a randomizing diffuser in a lensless imaging architecture, allowing us to recover close to the full spatial resolution of the sensor, which is not possible with traditional lens-based methods. Our method uses traditional pass-band spectral filters, but could be extended to photonic crystal slabs and other spectral filter designs. 

\textbf{Coded aperture methods} use a coded aperture, in combination with a dispersive optical element (e.g. a prism or diffractive grating), in order to modulate the light and encode spatial-spectral information~\cite{gehm2007single, lin2014spatial, wagadarikar2008single, cao2016computational}. These systems are able to capture hyperspectral images and videos but tend to be large table-top systems consisting of multiple lenses and optical components. In contrast, our system has a much smaller form factor, requiring only a camera sensor with an attached spectral filter array and a thin diffuser placed close to the sensor. 

\textbf{Speckle-based methods} use the wavelength dependence of speckle from a random media to achieve hyperspectral imaging. This has been demonstrated for compact spectrometers~\cite{redding2013compact, chakrabarti2015speckle} and extended to hyperspectral imaging~\cite{sahoo2017single, french2017speckle}. These systems can be compact, since they require only a sensor and scattering media as their optic; however their spectral resolution is limited by the speckle correlation through wavelengths. This is challenging to design for a given application, since the spatial and spectral resolutions are highly coupled. In contrast, our system uses spectral filters that can easily be adjusted for a given application and can be selected to have variable bandwidth or non-uniform spectral sampling.

\textbf{Dispersive methods} utilize the dispersion from a prism or diffractive optic to encode spectral information on the sensor. This can be accomplished opportunistically by a prism added to a standard DSLR camera~\cite{baek2017compact}. The resulting system has high spatial resolution, equal to that of the camera sensor, but spectral information is encoded only at the edges of objects in the scene, resulting in a highly ill-conditioned problem and lower spectral accuracy. Other methods use a diffuser (as opposed to a prism) as the dispersive element~\cite{golub2016compressed}. This can be more compact than prism-based systems and can have improved spatial resolution when combined with an additional RGB camera~\cite{hauser2020dual}. To further improve compactness,~\cite{jeon2019compact} uses a single diffractive optic as both the lens and the dispersive element, uniquely encoding spectral information in a spectrally-rotating point spread function (PSF). 

Our system uses a lensless architecture and a spectral filter array, together with sparsity assumptions, to reconstruct 3D hyperspectral information across 64 wavelengths. The design is most similar to~\cite{jeon2019compact} and achieves a similar compact size; however, our system achieves better spectral accuracy, and the use of the color filter array and diffuser results in more design flexibility, as our spectral and spatial resolutions are decoupled, enabling custom sensors tailored to specific spectral filter bands that do not need to be contiguous. 

\subsection{Lensless Imaging}
Lensless, mask-based imaging systems do not have a main lens, but instead use an amplitude or phase mask in place of imaging optics. These systems have been demonstrated for very compact, small form factor 2D imaging~\cite{asif2016flatcam, kuo2017diffusercam, tanida2001thin, tanida2003color}. They are generally amenable to compressive imaging, due to the multiplexing nature of lensless architectures; each point in the scene maps to many pixels on the sensor, allowing a sparse scene to be completely recovered from a subset of sensor pixels~\cite{fergus2006random}. Or, one can reconstruct higher-dimensional functions like 3D~\cite{antipa2018diffusercam} or video~\cite{antipa2019video} from a single 2D measurement. In this work, we use diffuser-based lensless imaging to spatially-multiplex light onto a repeated spectral filter array, then reconstruct 3D hyperspectral information. Because of the compressed sensing framework, our spatial resolution is better than the array super-pixel size, despite the missing information due to the array. 

\begin{figure}[t]
\centering
\includegraphics[width=\linewidth]{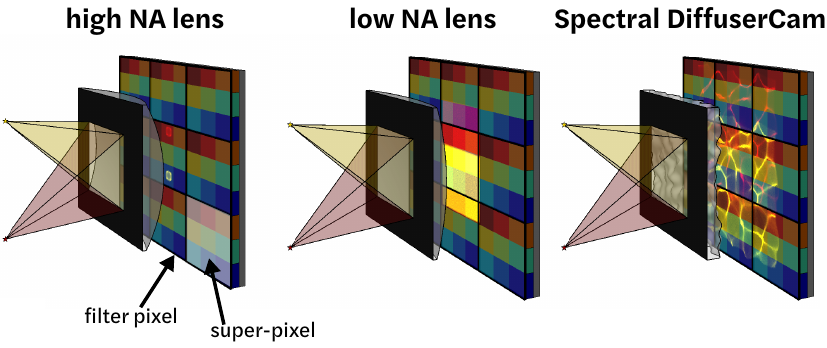}
\caption{Motivation for multiplexing: A high-NA lens captures high-resolution spatial information, but misses the yellow point source, since it comes into focus on a spectral filter pixel designed for blue light.  A low-NA lens blurs the image of each point source to be the size of the spectral filter's super-pixel, capturing accurate spectra at the cost of poor spatial resolution. Our DiffuserCam approach multiplexes the light from each point source across many super-pixels, enabling the computational recovery of both point sources and their spectra without sacrificing spatial resolution. Note that a simplified 3$\times$3 filter array is shown here for clarity.}
\label{fig:diffuser_vs_lens}
\end{figure}

\begin{figure}[tbh]
\centering
\includegraphics[width=\linewidth]{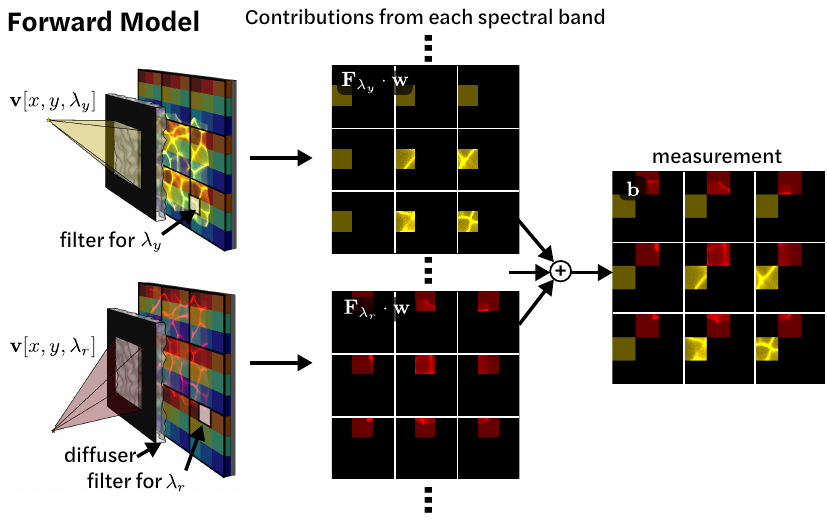}
\caption{Image formation model for a scene with two point sources of different colors, each with narrow-band irradiance centered at $\lambda_y$ (yellow) and $\lambda_r$ (red). The final measurement is the sum of the contributions from each individual spectral filter band in the array.  Due to the spatial multiplexing of the lensless architecture, all scene points $v(x,y,\lambda)$ project information to multiple spectral filters, which is why we can recover a high-resolution hyperspectral cube from a single image, after solving an inverse problem. }
%\caption{Image formation model for a scene with a single point source with two spectral peaks (a) and two point sources with overlapping spectra (b). Before going through the spectral filter, each point is convolved with the diffuser PSF. After passing through the spectral filter, each pattern is filtered by the filter function corresponding to the source wavelength. The captured image is the sum of the spectral bins. }
\label{fig:forward_model}
\end{figure}

\section{System Design Overview}
\label{sec:system_overview}
Our system leverages recent advances in both spectral filter array technology and compressive lensless imaging to decouple the spectral and spatial design. Furthermore, the spectral filter arrays can be deposited directly on the camera sensor. With a diffuser as our multiplexing optic, the system is compact and inexpensive at scale.

To motivate our need for a multiplexing optic instead of an imaging lens, let us consider three candidate architectures: one with a high numerical aperture (NA) lens whose diffraction-limited spot size is matched to the filter pixel size, one with a low-NA lens whose diffraction-limited spot size is matched to the super-pixel size, and finally our design with a diffuser as a multiplexing optic. Figure~\ref{fig:diffuser_vs_lens} illustrates these three scenarios with a simplified example of a spectral filter array consisting of $3\times3$ spectral filters (9 total) repeated horizontally and vertically.  Assume that the monochrome camera sensor has square pixels of lateral size $N_{\text{pixel}}$, the spectral filter array has square filters of size $N_{\text{filter}}$, and each $3\times 3$ block of spectral filters creates a \textit{super-pixel} of size $N_{\text{super-pixel}}$, where $N_{\text{pixel}} < N_{\text{filter}} < N_{\text{super-pixel}}$.

In the high-NA lens case, a point source in the scene will be imaged onto a single filter pixel of the sensor, and thus will only be measured if it is within the passband of that filter; otherwise it will not be recorded, Fig.~\ref{fig:diffuser_vs_lens} (left). In the low-NA lens case, each point source will be imaged to an area the size of the filter array super-pixel, and thus recorded by the sensor correctly, but at the price of low spatial-resolution (matched to the the super-pixel size), Fig.~\ref{fig:diffuser_vs_lens} (middle). In contrast, a multiplexing optic can avoid the gaps in the measurement of the high-NA lens and achieve better resolution than the low-NA case.

A diffuser multiplexes the light from each point source such that it hits many filter pixels, covering all of the spectral bands. And the spatial resolution of the final image can be on the order of the camera pixel size, provided that conditions for compressed sensing are met, Fig.~\ref{fig:diffuser_vs_lens} (right). In practice, the spatial resolution of our system will be bounded by the autocorrelation of the point spread function (PSF), as detailed in Sec.~\ref{sec:system_analysis}, and the diffuser PSF must span multiple super-pixels to ensure that each point in the world is captured. Since compressive recovery is used to recover a 3D hyperspectral cube from a 2D measurement, the resolution is a function of the scene complexity, as described in Sec.~\ref{sec:system_analysis}. 

%the diffuser PSF must span multiple super-pixels to ensure that each point in the world is captured however the SNR decreases as the PSF size increases since the light is spread across more pixels.  

\section{Imaging Forward Model}
Given our design with a diffuser placed in front of a sensor that has a spectral filter array on top of it, in this section we outline a forward model for the optical system, illustrated in Fig.~\ref{fig:forward_model}. This model is a critical piece of our iterative inverse algorithm for hyperspectral reconstruction and will also be used to analyze spatial and spectral resolution.

%\the\textwidth

\subsection{Spectral filter model}

%The spectral filter array is placed on top of an imaging sensor, such that the exposure at each point on the sensor, $L(x,y)$, can be modeled as a spectral integral,

%\begin{equation}
% L(x,y) = \int_{\lambda_1}^{\lambda_2} F(\lambda |x,y) \cdot v(x,y,\lambda) d\lambda,
%\end{equation}

%\noindent where $\cdot$ denotes point-wise multiplication, $v(x,y,\lambda)$ is the spectral irradiance incident on the filter array and $F(\lambda |x,y)$ is a 3D continuous function describing the transmittance of light through the spectral filter, which we call the \textit{filter function}. In this model, we absorb the sensor's spectral response into the definition of $F(\lambda |x,y)$. Our device's filter function is determined experimentally (see Sec~\ref{sec:inplementation}.C)  and shown in Fig.~\ref{fig:spectral_respose}(b).This can be generalized to any arbitrary spectral filter design and does not assume alignment between the filter pixels and the sensor pixels. Here, we focus on the case of a repeating grid of spectral filters, where each 'super-pixel' consists of a set of narrow-band filters. Our device has a 8$\times$8 grid of filters in each super-pixel; Fig.~\ref{fig:forward_model} illustrates a simplified 3$\times$3 grid, for clarity.  

The spectral filter array is placed on top of an imaging sensor, such that the exposure on each pixel is the sum of point-wise multiplications with the discrete filter function,

\begin{equation}
 \mathbf L[x,y] = \sum_{\lambda=0}^{K-1} \mathbf F_{\lambda}[x,y] \cdot \mathbf v[x,y,\lambda] ,
\end{equation}

\noindent where $\cdot$ denotes point-wise multiplication, $\textbf v[x,y,\lambda]$ is the spectral irradiance incident on the filter array and $\textbf F_{\lambda}[x,y]$ is a 3D function describing the transmittance of light through the spectral filter for $K$ wavelength bands, which we call the \textit{filter function}. In this model, we absorb the sensor's spectral response into the definition of $\textbf F_{\lambda}[x,y]$. Our device's filter function is determined experimentally (see Sec~\ref{sec:inplementation}.C)  and shown in Fig.~\ref{fig:spectral_respose}(b). This can be generalized to any arbitrary spectral filter design and does not assume alignment between the filter pixels and the sensor pixels. Here, we focus on the case of a repeating grid of spectral filters, where each 'super-pixel' consists of a set of narrow-band filters. Our device has a 8$\times$8 grid of filters in each super-pixel; Fig.~\ref{fig:forward_model} illustrates a simplified 3$\times$3 grid, for clarity. 

%\noindent where $x$ and $y$ are the sensor rows and columns and $\textbf F_{\lambda}[x,y] = F(\lambda |x,y)$ is the discrete filter function at wavelength $\lambda$. There are a total of $K$ wavelengths.  

\subsection{Diffuser model}
The diffuser (a smooth pseudorandom phase optic) in our system achieves spatial multiplexing; this results in a compact form factor and enables reconstruction with spatial resolution better than the super-pixel size via compressed sensing. The diffuser is placed a small distance away from the sensor and an aperture is placed on the diffuser to limit higher angles. The sensor plane intensity resulting from the diffuser can be modeled as a convolution of the scene, $\textbf v[x,y,\lambda]$ with the on-axis PSF, $\textbf h[x,y]$~\cite{kuo2017diffusercam}:

%\begin{equation}
% w(x,y, \lambda) = v(x,y,\lambda) \stackrel{(x,y)}{*} h(x,y)
%\end{equation}

%\noindent where $\stackrel{(x,y)}{*}$ represents a 2D convolution over spatial dimensions. In discrete pixel coordinates, this can be denoted as:

\begin{equation}
 \mathbf{w}[x,y,\lambda] = \text{crop} \Big( \mathbf v[x,y,\lambda] \stackrel{[x,y]}{*} \mathbf h[x,y]) \Big)
\end{equation}

\noindent where $\stackrel{[x,y]}{*}$ represents a discrete 2D linear convolution over spatial dimensions. The crop function accounts for the finite sensor size. We assume that the PSF does not vary with wavelength and validate this experimentally in Sec.~\ref{sec:inplementation}.B. However, this model can be easily extended to include a spectrally-varying PSF, $\mathbf h[x,y,\lambda]$ if there is more dispersion across wavelengths.
%\the\linewidth

We assume that objects are placed beyond the hyperfocal distance of the imager, therefore the PSF has negligible depth-variance and a 2D convolutional model is valid~\cite{kuo2017diffusercam}. If objects are placed within the hyperfocal distance, a 3D model will be needed to account for the depth-variance of the PSF.

\subsection{Combined model}
Combining the spectral filter model with the diffuser model, we have the following discrete forward model:

\begin{align}
\label{eqn:forward_combined}
 \mathbf b &= \sum_{\lambda=0}^{K-1} \mathbf F_{\lambda}[x,y] \cdot \text{crop} \Big( \mathbf h[x,y] \stackrel{[x,y]}{*} \mathbf v[x,y,\lambda] \Big) \\
 &= \sum_{\lambda=0}^{K-1} \mathbf F_{\lambda}[x,y] \cdot \mathbf{w}[x,y,\lambda] \\
 &= \mathbf A \mathbf v.
\end{align}

\noindent The linear forward model is represented by the combined operations in matrix $\mathbf A$. Figure~\ref{fig:forward_model} illustrates the forward model for several point sources, showing the intermediate variable $\mathbf{w}[x,y,\lambda]$, which is the scene convolved with the PSF, before point-wise multiplication by the filter function. The final image is the sum over all wavelengths. 

\section{Hyperspectral Reconstruction}
To recover the hyperspectral datacube from the 2D measurement, we must solve an underdetermined inverse problem. Since our system falls within the framework of compressive sensing due to our incoherent, multiplexed measurement, we use $l_1$ minimization.  We use a weighted 3D total variation (3DTV) prior on the scene, as well as a non-negativity constraint, and a low-rank prior on the spectrum.  This can be written as:

\begin{equation}
\label{eqn:inverse}
 \hat{\mathbf v} = \argmin_{\mathbf v\geq 0} \frac{1}{2} \| \mathbf b - \mathbf A \mathbf v \|_2^2 + \tau_1 \| \nabla_{xy\lambda} \mathbf v\|_1 + \tau_2\|\mathbf v\|_*,
\end{equation}

\noindent where $\nabla_{xy\lambda} = [\nabla_x \nabla_y \nabla_\lambda]^T$ is the matrix of forward finite differences in the $x$, $y$, and $\lambda$ directions, $\|\cdot \|_*$ represents the nuclear norm, which is the sum of singular values. $\tau_1$ and $\tau_2$ are the tuning parameters for the 3DTV prior and low-rank priors, respectively.  We use the fast iterative shrinkage-thresholding algorithm (FISTA)~\cite{beck2009fast} with weighted anisotropic 3DTV to solve this problem according to~\cite{kamilov2016parallel}.

\begin{figure}[t!]
\centering
\includegraphics[width=\linewidth]{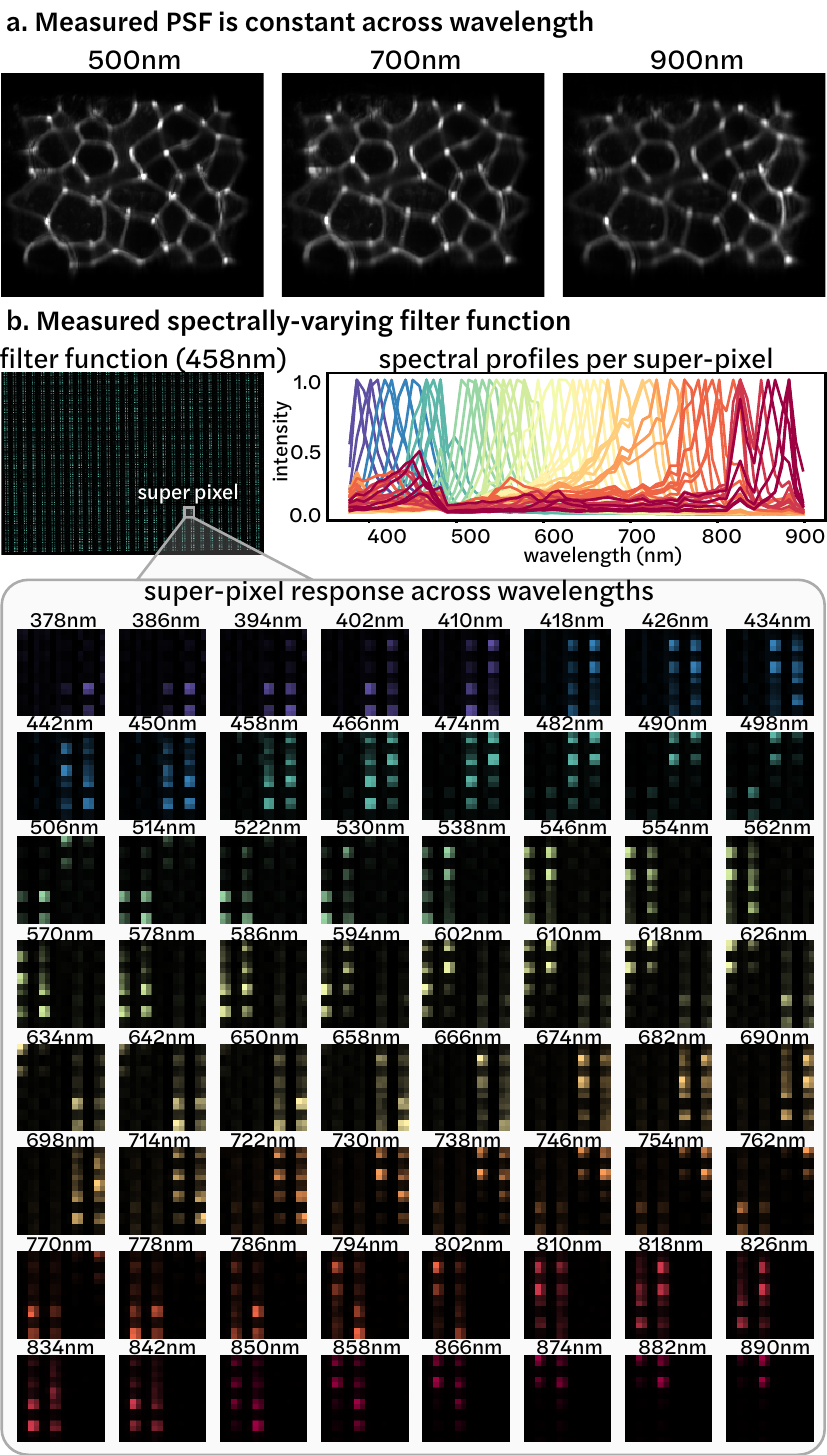}
\caption{Experimental calibration of Spectral DiffuserCam. (a) The caustic PSF (contrast-stretched and cropped), before passing through the spectral filter array, is similar at all wavelengths. (b) The spectral response with the filter array only (no diffuser). (Top left) Full measurement with illumination by a 458nm plane wave. The filter array consists of 8$\times$8 grids of spectral filters repeating in 28$\times$20 super-pixels. (Top right) Spectral responses of each of the 64 color channels. (Bottom) Spectral response of a single super-pixel as illumination wavelength is varied with a monochromater. }
\label{fig:spectral_respose}
\end{figure}

\section{Implementation Details}
\label{sec:inplementation}
%\subsection{Prototype}
We built a prototype system using a CMOS sensor, a hyperspectral filter array provided by Viavi Solutions (Santa Rosa, CA)\cite{saxe2018advances}, and an off-the-shelf diffuser (Luminit 0.5°) placed 1cm away from the sensor. The sensor has 659$\times$494 pixels (with a pixel pitch of 9.9$\mu m$), which we crop down to 448$\times$320 to match the spectral filter array size. The spectral filter array consists of a grid of 28$\times$20 super-pixels, each with an 8$\times$8 grid of filter pixels (64 total, spanning the range 386-898nm). Each filter pixel is 20$\mu m$ in size, covering slightly more than 4 sensor pixels. The alignment between the sensor pixels and the filter pixels is unknown, requiring a calibration procedure (detailed in Sec.~\ref{sec:inplementation}\ref{ssec:filterfunctioncalibration}). The exposure time is adjusted for each image, ranging from 1ms-13ms, which is short enough for video-rate acquisition. The computational reconstruction typically takes 12-24 minutes (for 500-1000 iterations) on an RTX 2080-Ti GPU using MATLAB. 

% sensor: 101 lp/mm
% nyquist 50.5 lp/mm

\subsection{Filter Function Calibration}
\label{ssec:filterfunctioncalibration}
To calibrate the filter function ($\mathbf F_\lambda [x,y]$ in Eqn.~\ref{eqn:forward_combined}), including the spectral sensitivity of both the sensor and the spectral filter array, we use a Cornerstone 130 1/3m motorized monochromator (Model 74004). The monochromater creates a narrow-band source of 5nm full-width at half-maximum (FWHM) and we measure the filter response (without the diffuser) while sweeping the source by 8nm increments from 386nm to 898nm. The result is shown in Fig.~\ref{fig:spectral_respose}(b). 

\subsection{PSF Calibration}
We also need to calibrate the diffuser response by measuring the diffuser PSF pattern without the spectral filter array. Because the diffuser is relatively smooth with large features (relative to the wavelength of light), the PSF remains relatively constant as a function of wavelength, as shown in Fig.~\ref{fig:spectral_respose}(a). Hence, we only need to calibrate for a single wavelength by capturing a single point source calibration image~\cite{antipa2018diffusercam}. However, this is not trivial because the spectral filter array is bonded to the sensor and cannot be removed easily. In our setup, we instead take advantage of the fact that our filter array is smaller than our sensor, so we can measure the PSF using the edges of the raw sensor, by shifting the point source to scan the different parts of the PSF over the raw sensor area and stitching the sub-images together. In a system where the filter size is matched to the sensor, this trick will not be possible, but an optimization-based approach could be developed to recover the PSF from measurements. 
%This is not be possible for systems in which the filter array is matched to the sensor size, so we also provide and validate an optimization-based approach that recovers the PSF from several randomly shifted measurement images of a point source with known spectra. See Supplement for details.

\subsection{System Non-idealities}
Our reconstruction quality and spectral resolution are limited by two non-idealities in our system. First, our camera development board performs an undefined and uncontrollable non-linear contrast-stretching to all images.  This makes the measurement non-linear and impedes our imaging of dim objects (since the camera performs a larger contrast stretching for dimmer images). Further, our spectral calibration may have errors, since each calibration image cannot be normalized by the intensity of light hitting the sensor. This may cause certain wavelength bands to appear brighter or dimmer than they should in our spectral reconstructions. A better camera board without automatic contrast stretching should fix this problem and provide more quantitative spectral profile reconstructions in the future.

Second, we used a simplified spectral calibration in which we measured the response with uniform spectral sampling, instead of at the true wavelengths of the filters.  Due to the mismatch between our calibration scheme (measured every 8nm with constant bandwidth) and the actual spectral filters (center wavelengths spaced 5-12nm apart with bandwidths between 6-23nm), sometimes our calibration wavelengths fall between two filters, resulting in an ambiguity.  Given this non-ideal calibration, our effective spectral bands are limited to 49 bands, instead of 64. In our results, we show all 64 bands, but note that some will have overlapping spectral responses. In the future, we will calibrate at the design wavelengths of the filter to fix this issue. Further, the deposition of the spectral filters directly on-top of the camera pixels (requiring precise placement during the manufacturing stage) would alleviate the need for this calibration entirely.

\section{Resolution Analysis}
\label{sec:system_analysis}
Here, we derive our theoretical resolution and experimentally validate it with our prototype system. First, we discuss spectral resolution, which is set by the filter bandwidths, and then we compute the expected two-point spatial resolution, based on the PSF autocorrelation. Since our resolution is scene-dependent, we expect the resolution to degrade with scene complexity. To characterize this, we present theory for multi-point resolution based on the condition number analysis introduced in~\cite{antipa2018diffusercam}. We compare our system against those with a high-NA and low-NA lens instead of a diffuser. Our results demonstrate two-point spatial resolution of \mytilde{}0.19 super-pixels and multi-point spatial resolution of \mytilde{}0.3 super-pixels for 64 spectral channels ranging from 386-898nm. 

%and  quantify our system resolution and compare our system against several baselines that use a lens instead of a diffuser. In order to Nyquist match the super-pixels of the filter array, a lens should have a diffraction-limited spot size large enough to cover the full super-pixel (160$\mu m$), however this corresponds to limiting the resolution of the system by 8$\times$ from that of sensor resolution. We compare against systems consisting of a lens with a diffraction-limited spot size covering a single spectral filter (20$\mu m$, high NA lens), and covering a full super-pixel (160$\mu m$, low NA lens). 

\begin{figure}[h]
\centering
\includegraphics[width=\linewidth]{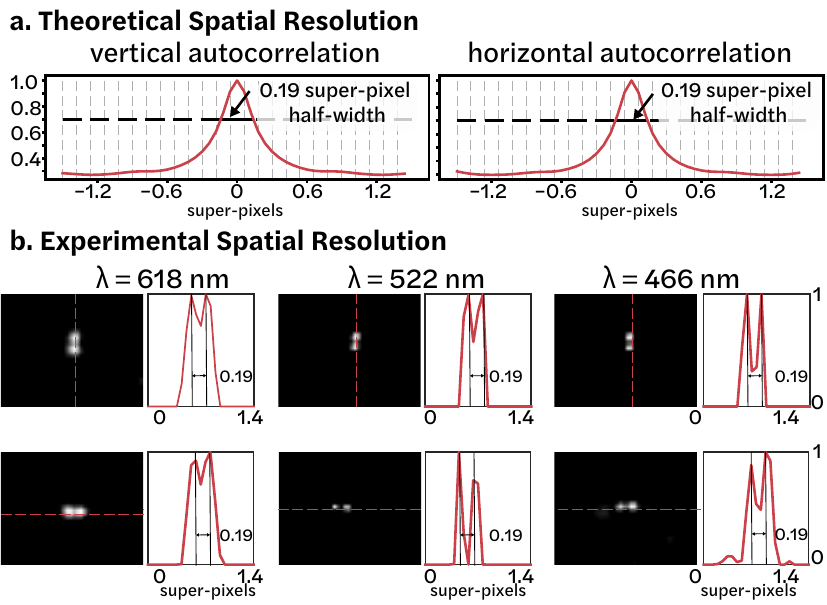}
\caption{Spatial Resolution analysis. (a) The theoretical resolution of our system, defined as the half-width of the autocorrelation peak at 70$\%$ its maximum value, is 0.19 super-pixels. (b) Experimental two-point reconstructions demonstrate 0.19 super-pixel resolution across all wavelengths (slices of the reconstruction shown here), matching the theoretical resolution.}
\label{fig:spatial_res}
\end{figure}

%First, we present our theoretical two point spatial resolution and spectral resolutions and validate these experimentally. Then, we present simulation results comparing the spatial resolution of our system to that of different lens-based systems with varying diffraction limited spot-sizes. Finally, we present a condition number analysis to predict our resolution for varying scene complexity.

\subsection{Spectral Resolution}
\begin{figure}[t]
\centering
\includegraphics[width=\linewidth]{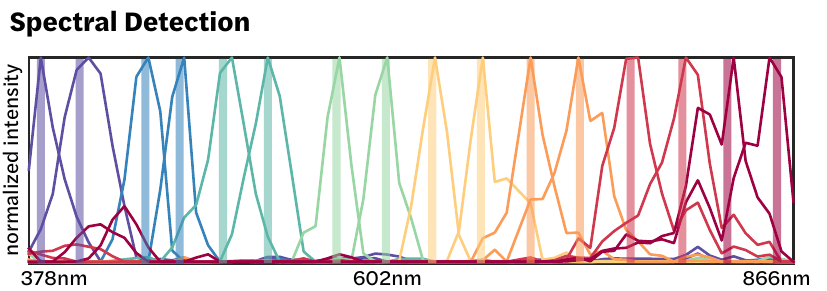}
\caption{Spectral resolution analysis. Sample spectra from hyperspectral reconstructions of narrow-band point sources, overlaid on top of each other, with shaded lines indicating the ground-truth. For each case, the recovered spectral peak matches the true wavelength within 5nm.} %(b) Two-point spectral resolution varies from 23 nm to 46 nm, as determined by applying Rayleigh's criterion to a reconstruction of synthetically added point sources with different wavelengths.}
\label{fig:spectral_res}
\end{figure}

%\subsubsection{Spectral resolution}
Spectral resolution is determined by the spectral channels of the filter array. As such, we expect to be able to resolve the 64 spectral channels present in our spectral filter array.  The filters have an average spacing of 8nm across a 386-898nm range with bandwidths between 6-23nm. To validate our spectral resolution, we scan a point source across those wavelengths using a monochrometer. Figure~\ref{fig:spectral_res} shows a sampling of spectral reconstructions overlaid on top of each other, with the shaded blocks indicating the ground-truth monochrometer spectra. Our reconstructions all match the ground-truth peaks within 5nm of the true wavelength. The small red peaks around 400nm are artifacts from the monochrometer, which emitted a 2nd peak around 400nm for the longer wavelengths. %To determine the two-point spectral resolution of our system, we synthetically add the raw data from point sources at two different wavelengths and reconstruct a scene of two points at the same spatial position with varying separations in wavelength. We determine the spectral resolution by applying the Rayleigh criterion to the spectral dimension of the reconstruction. Figure~\ref{fig:spectral_res}(b) shows the measured spectral resolution and how it varies with wavelength. Our system achieves 30nm two-point spectral resolution across most of the 390-900nm spectral range.

\subsection{Two-point Spatial Resolution}
%\subsubsection{Spatial resolution}
Spatial resolution of our system, in terms of the two-point resolution, will be bounded by that of a lensless imager with the diffuser only (without the spectral filter array). The expected resolution can be defined as the autocorrelation peak half-width at 70\% the maximum value~\cite{kuo2017diffusercam}, Fig.~\ref{fig:spatial_res}(a). For our system, this is \mytilde{}3 sensor pixels, or 0.19 super-pixels. To experimentally measure the spatial resolution of our system, we image two point sources at three different wavelengths ($618$ nm, $522$ nm, $466$ nm).  The reconstructions in Fig.~\ref{fig:spatial_res} show that we can resolve two point sources that are 0.19 super-pixels apart for each wavelength and orientation, as determined by applying the Rayleigh criterion. This demonstrates that our system achieves sub-super-pixel spatial resolution, consistent with the expected resolution that would be achieved without the spectral filter array. 

%To determine the spatial resolution of our system and how it varies with wavelengths, we image two point sources at three different wavelengths ($626$ nm, $522$ nm, $456$ nm). Fig.~\ref{fig:spatial_res}{a} shows the expected lateral resolution based on applying the Rayleigh criterion to the autocorrelation of the PSF (~3 pixels). The reconstructions in Fig.~\ref{fig:spatial_res} show the we can resolve two point sources $3$-$4$ pixels apart. This demonstrates that using a diffuser achieves sub-superpixel spatial resolution and matches our expected lateral resolution.

\begin{figure}[h]
\centering
\includegraphics[width=\linewidth]{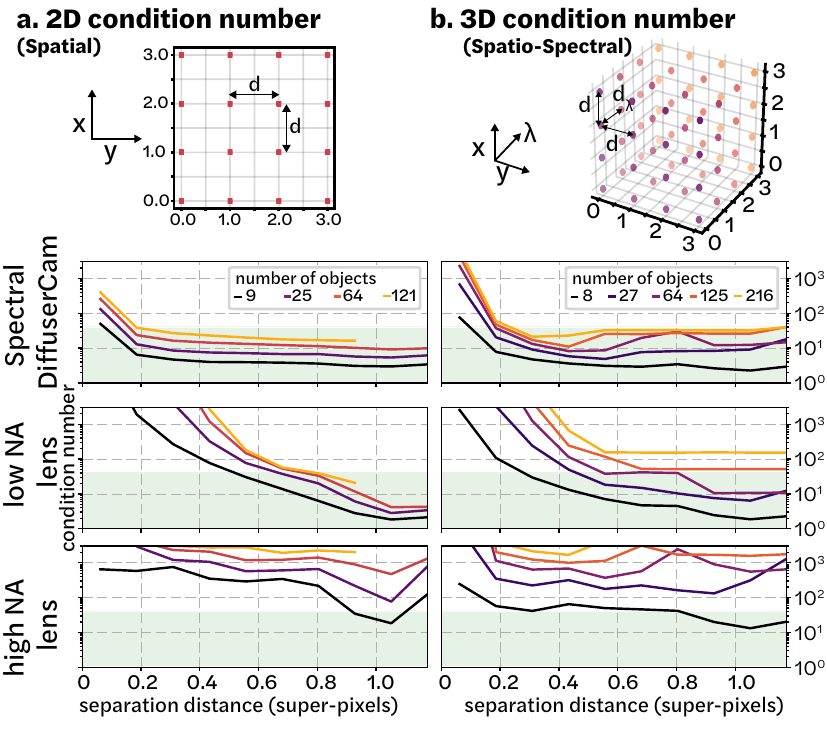}
\caption{Condition number analysis for Spectral DiffuserCam, as compared to a low-NA or high-NA lens. (a) Condition numbers for the 2D spatial case (single spectral channel) are calculated by generating different numbers of points on a 2D grid, each with separation distance $d$. (b) Condition numbers for the full spatio-spectral case are calculated on a 3D grid. A condition number below 40 is considered to be good (shown in green). The diffuser has a consistently better performance for small separation distances than either the low-NA or the high-NA lens. The diffuser can resolve objects as low as 0.3 super-pixels apart for more complex scenes, whereas the low-NA lens requires larger separation distances  and the high-NA lens suffers errors due to gaps in the measurement. }
\label{fig:cond_num}
\end{figure}

\subsection{Multi-point resolution}
%\subsubsection{Local condition number analysis}
Because our image reconstruction algorithm contains nonlinear regularization terms, our reconstruction resolution will be object dependent. Hence, two-point resolution measurements are not sufficient for fully characterizing the system resolution, and should be considered a \textit{best case} scenario. To better predict real-world performance, we perform a local condition number analysis, as introduced in~\cite{antipa2018diffusercam}, that estimates resolution as a function of object complexity. The local condition number is a proxy for how well the forward model can be inverted, given known support, and is useful for systems such as ours in which the full $\mathbf A$ matrix is never explicitly calculated~\cite{candes2014towards}.  

The local condition number theory states that given knowledge of the \textit{a priori} support of the scene, $\mathbf v$, we can form a sub-matrix consisting only of columns of $\mathbf A$ corresponding to the non-zero voxels. The reconstruction problem will be ill-posed if any of the sub-matrices of $\mathbf A$ are ill-conditioned, which can be quantified by the condition number of the sub-matrices. The worst-case condition number will be when sources are near each other, therefore we compute the condition number for a group of point sources with a separation varying by an integer number of voxels and repeat this for increasing numbers of point sources. 

\begin{figure}[tbh!]
\centering
\includegraphics[width=\linewidth]{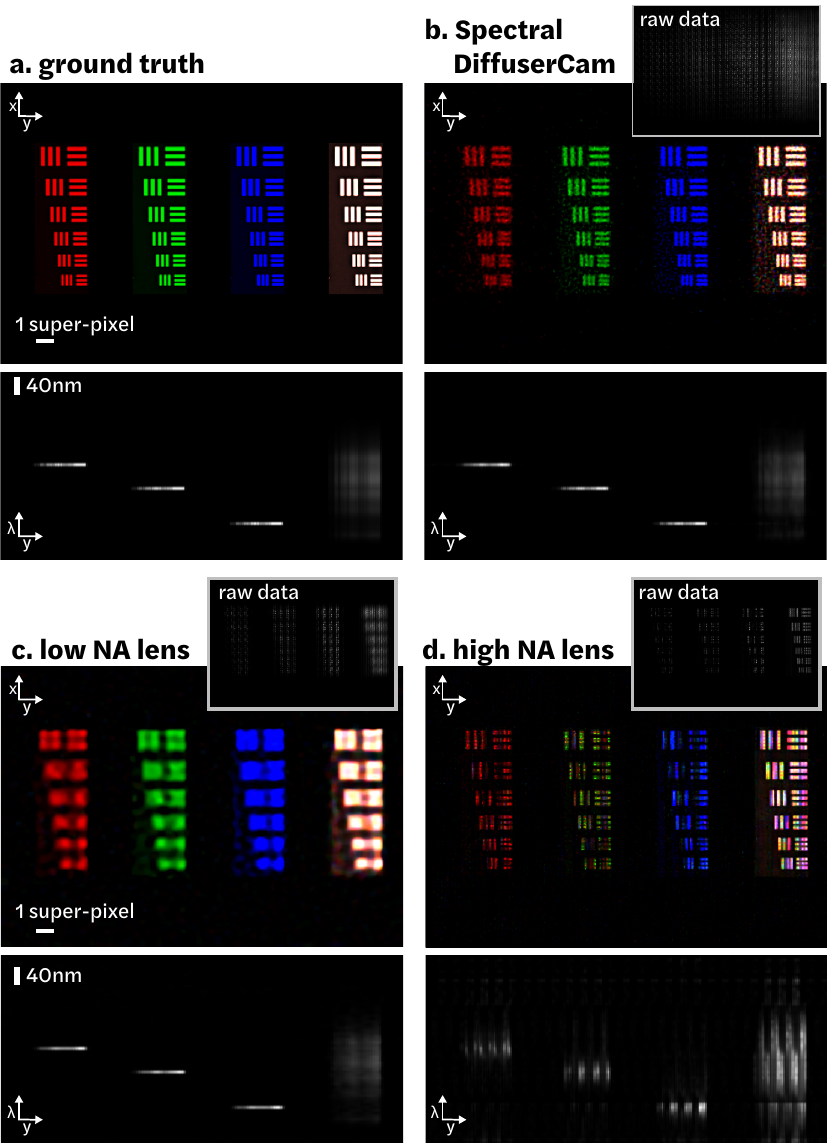}
\caption{Simulated hyperspectral reconstructions comparing our Spectral DiffuserCam result with alternative design options. (a) Resolution target with different sections illuminated by narrow-band 634nm (red), 570nm (green), 474nm (blue), and broadband (white) sources. (b) Reconstruction of the target by Spectral DiffuserCam, (c) a low-NA lens design, and (d) a high-NA lens design, each showing the raw data, false-colored reconstruction and $\lambda y$ sum projection. The diffuser achieves higher spatial resolution and better accuracy than the low-NA and the high-NA lens.
%, and has lower MSE for higher frequency content.
%, showing a lower error at the edges of the resolution target than the low NA lens
%(b) Reconstruction of a resolution target illuminated by a broadband source: (top) $xy$ slice at wavelength 586nm, (middle) $y\lambda$ sum projection, (bottom) the MSE error. In both cases, the diffuser achieves higher spatial resolution than either the low NA lens or the high NA lens, and has lower MSE for higher frequency content.%, showing a lower error at the edges of the resolution target than the low NA lens.
}
\label{fig:res_target}
\end{figure}

In Fig.~\ref{fig:cond_num}, we calculate the local condition number for two cases: the 2D spatial reconstruction case, considering only a single spectral channel, and the 3D case, considering points with varying spatial and spectral positions. For comparison, we also simulate the condition number for a low-NA and high-NA lens, as introduced in Sec.~\ref{sec:system_overview}. The results show that our diffuser design has a consistently lower condition number than either the low- or high-NA lens, having a condition number below 40 for separation distances of greater than \mytilde{}0.3 super-pixels. The low-NA lens needs a separation distance closer to \mytilde{}1 super-pixel, as expected, and the high-NA lens has an erratic condition number due to the missing information in the measurement. 

From this analysis, we can see that, beyond 0.3 super-pixels separation, the condition number for the diffuser does not get arbitrarily worse for increasing scene complexity. Thus, our expected spatial resolution is approximately 0.3 super-pixels.

\subsection{Simulated Resolution Target Reconstruction}
Next, we validate the results of our condition number analysis through simulated reconstructions of a resolution target with different spatial locations illuminated by different sources (red, green, blue and white light), as shown in Fig.~\ref{fig:res_target}. For each simulation, we add Gaussian noise with a variance of \num{1e-5} and run the reconstruction for 2,000 iterations of FISTA with 3DTV. Our system resolves features that are 0.3 super-pixels apart, whereas the low-NA lens can only resolve features that are roughly 1 super-pixel apart and the high-NA lens results in gaps, validating our predicted performance. 
%The spectral error map confirms that the diffuser is able to accurately recover higher spatio-spectral resolution than the low-NA lens, which has most of the spectral error concentrated at the edges of the reconstruction target.  

\begin{figure}[h]
\centering
\includegraphics[width=\linewidth]{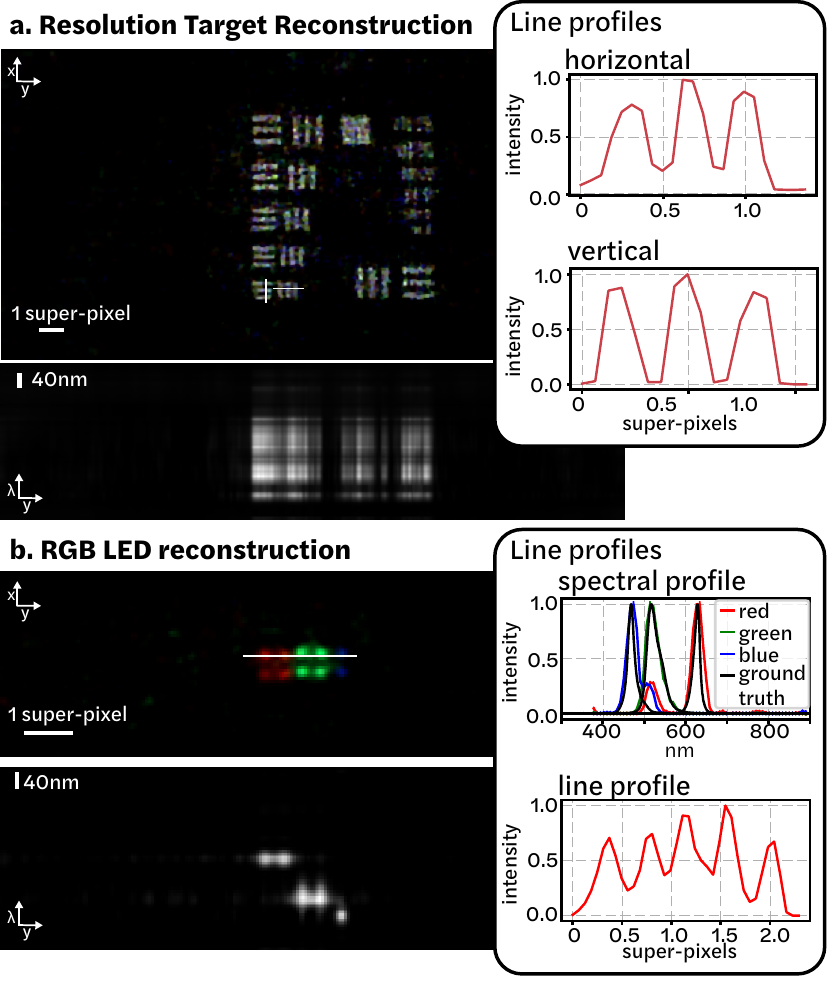}
\caption{(a) Experimental reconstruction of a broadband resolution target, showing the $xy$ sum projection (top) and $\lambda y$ sum projection (bottom), demonstrating spatial resolution of 0.3 super-pixels.  (b) Experimental reconstruction of 10 multi-colored LEDs in a grid with \mytilde{}0.4 super-pixels spacing (four red LEDs on left, four green in middle, two blue at right).  We show the $xy$ sum projection (top) and $\lambda y$ sum projection (bottom). The LEDs are clearly resolved spatially and spectrally, and spectral line profiles for each color LED closely match the ground truth spectra from a spectrometer. }
\label{fig:res_target_exp}
\end{figure}

\section{Experimental Results}
We start with experimental reconstructions of simple objects with known properties - a broadband USAF resolution target displayed on a computer monitor, and a grid of RGB LEDs (Fig.~\ref{fig:res_target_exp}). We resolve points that are \mytilde{}.3 super-pixels apart, which matches our expected multi-point resolution based on the condition number analysis above. For the RGB LED scene, the ground truth spectral profiles of the LEDs are measured using a spectrometer, and our recovered spectral profile closely matches the ground truth, as shown in Fig.~\ref{fig:res_target_exp}(b).

\begin{figure*}[t]
\centering
\includegraphics[width=\linewidth]{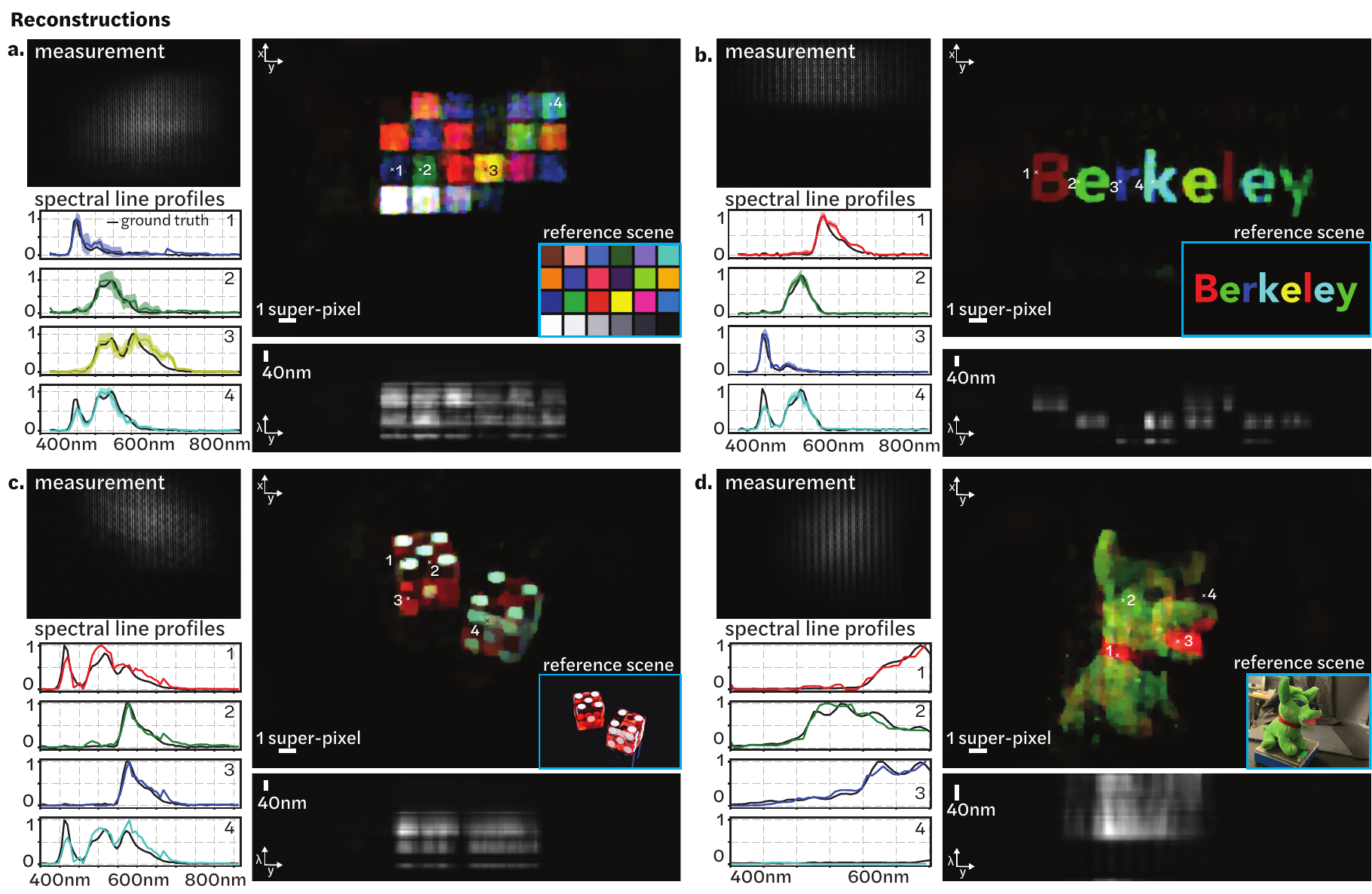}
\caption{Experimental hyperspectral reconstructions.  (a-c) Reconstructions of color images displayed on a computer monitor and (d) Thorlabs plush toy placed in front of the imager and illuminated by two Halogen lamps. The raw measurement, false color images, $x\lambda$ sum projections and spectral line profiles for four spatial points are shown for each scene. The ground truth spectral line profiles, measured using a spectrometer, are plotted in black for reference. Spectral line profiles in (a,b) show the average and standard deviation spectral profiles across the area of the box or letter in the object, whereas (c-d) show a line profile from a single spatial point in the scene. }
\label{fig:results}
\end{figure*}

Next, we show reconstructions of more complex objects, either displayed on a computer monitor or illuminated with two halogen lamps (Figure~\ref{fig:results}). We plot the ground truth spectral line profiles, as measured by a Thorlabs CCS200 spectrometer, from four points in the scene, showing that we can accurately recover the spectra. A reference RGB scene is shown for each image, demonstrating that the reconstructions spatially match the expected scene.   %We recover a 3D hyperspectral cube with 64 spectral channels from the raw grayscale measurement. 

\section{Discussion}
A key advantage of our design over previous work is its flexibility to choose the spectral filters in order to tailor the system to a specific application. For example, one can non-linearly sample a wide range of wavelengths (which is difficult with many previous snapshot hyperspectral imagers). In future, we plan to design implementations specific to various task-based applications, which could make hyperspectral imaging more easily adopted, especially since the price is several orders-of-magnitude lower than currently available hyperspectral cameras. 

Currently, we experimentally achieve a spatial resolution of \mytilde{}0.3 super-pixels, or 5 sensor pixels.  In future designs, we should be able to achieve the full sensor resolution (along with better quality reconstructions) by optimizing the randomizing optic, instead of using an off-the shelf diffuser. This could be achieved by end-to-end optical design~\cite{sitzmann2018end, peng2019learned}.

Our system has two main limitations: light-throughput and scene-dependence.  Due to the use of narrow-band spectral filters, much of the light is filtered out by the filters. This provides good spectral accuracy and discrimination, but at the cost of low light throughput. In addition, since the light is spread by the diffuser over many pixels, the signal-to-noise ratio (SNR) is further decreased. Hence, our imager is not currently suitable for low-light conditions. This light-throughput limitation can be mitigated in the future by the use of photonic crystal slabs instead of narrowband filters, in order to increase light-throughput while maintaining spatio-spectral resolution and accuracy~\cite{wang2019single}. In addition, end-to-end design of both the spectral filters and the phase mask should improve efficiency, since application-specific designs can use only the set of wavelengths necessary for a particular task, without sampling the in-between wavelengths. Reducing the number of spectral bands improves both light throughput (because more sensor area will be dedicated to each spectral band) and spatial resolution (because the super-pixels will be smaller).  

Our second limitation is scene-dependence, as our reconstruction algorithm relies on object sparsity (e.g. sparse gradients). Because of the non-linear regularization term, it is difficult to predict performance, and one might suffer artifacts if the scene is not sufficiently sparse. Recent advances in machine learning and inverse problems seek to provide better signal representations, enabling the reconstruction of more complicated, denser scenes~\cite{liu2020information, bora2017compressed}.  In addition, machine learning could be useful in speeding up the reconstruction algorithm~\cite{monakhova2019learned} as well as potentially utilizing the imager more directly for a higher-level task, such as classification~\cite{diamond2017dirty}.

\section{Conclusion}
Our work presents a new hyperspectral imaging modality that combines a color filter array and lensless imaging techniques for an ultra-compact and inexpensive hyperspectral camera.  The spectral filter array encodes spectral information onto the sensor and the diffuser multiplexes the incoming light such that each point in the world maps to many spectral filters. The multiplexed nature of the measurement allows us to use compressive sensing to reconstruct high spatio-spectral resolution from a single 2D measurement. We provided an analysis for the expected resolution of our imager and experimentally characterized the two-point and multi-point resolution of the system. Finally, we built a prototype and demonstrated reconstructions of complex spatio-spectral scenes, achieving up to 0.19 super-pixel spatial resolution across 64 spectral bands.

\section*{Funding Information}
This work was supported by the Gordon and Betty Moore Foundation Data-Driven Discovery Initiative Grant GBMF4562, and STROBE: A National Science Foundation Science \& Technology Center under Grant No. DMR 1548924. Kristina Monakhova and Kyrollos Yanny acknowledge funding from the National Science Foundation Graduate Research Fellowship Program (NSF GRFP) (DGE 1752814). The camera and spectral filter array were provided by Viavi Solutions (Santa Rosa, CA).

\section*{Acknowledgments}
The authors would like to thank Viavi Solutions (Santa Rosa, CA), and particularly Bill Houck, for their technical help and support, as well as Nick Antipa and Grace Kuo for helpful discussions.

\section*{Disclosures}
The authors declare no conflicts of interest.
% \newline
% See Supplement 1 for supporting content.

% Bibliography

\bibliography{sample}
\bibliographyfullrefs{sample}

\end{document}